\documentclass[twocolumn,tighten]{aastex62}

\received{}
\revised{}
\accepted{}

\shorttitle{Lyman Break Radio Galaxy at z = 4.7}
\shortauthors{Yamashita et al.}

\begin{document}

\title{Wide and Deep Exploration of Radio Galaxies with Subaru HSC (WERGS). III.\\
Discovery of a $z = 4.72$ Radio Galaxy with Lyman Break Technique}

\correspondingauthor{Takuji Yamashita}

\author{Takuji Yamashita}
\affil{National Astronomical Observatory of Japan, 2-21-1 Osawa, Mitaka, Tokyo 181-8588, Japan}
\affiliation{Research Center for Space and Cosmic Evolution, Ehime University, 2-5 Bunkyo-cho, Matsuyama, Ehime 790-8577, Japan}
\email{takuji.yamashita@nao.ac.jp}

\author{Tohru Nagao}
\affiliation{Research Center for Space and Cosmic Evolution, Ehime University, 2-5 Bunkyo-cho, Matsuyama, Ehime 790-8577, Japan}

\author{Hiroyuki Ikeda}
\affiliation{National Astronomical Observatory of Japan, 2-21-1 Osawa, Mitaka, Tokyo 181-8588, Japan}
\affiliation{National Institute of Technology, Wakayama College, 77 Noshima, Nada, Gobo, Wakayama 644-0023, Japan}

\author{Yoshiki Toba}
\affiliation{Department of Astronomy, Kyoto University, Kitashirakawa-Oiwake-cho, Sakyo-ku, Kyoto 606-8502, Japan}
\affiliation{Academia Sinica Institute of Astronomy and Astrophysics, 11F of Astronomy-Mathematics Building, AS/NTU, No.1, Section 4,
Roosevelt Road, Taipei 10617, Taiwan}
\affiliation{Research Center for Space and Cosmic Evolution, Ehime University, 2-5 Bunkyo-cho, Matsuyama, Ehime 790-8577, Japan}

\author{Masaru Kajisawa}
\affiliation{Research Center for Space and Cosmic Evolution, Ehime University, 2-5 Bunkyo-cho, Matsuyama, Ehime 790-8577, Japan}
\affiliation{Graduate School of Science and Engineering, Ehime University, Bunkyo-cho, Matsuyama, Ehime 790-8577, Japan}

\author{Yoshiaki Ono}
\affiliation{Institute for Cosmic Ray Research, The University of Tokyo, 5-1-5 Kashiwanoha, Kashiwa, Chiba 277-8582, Japan}

\author{Masayuki Tanaka}
\affiliation{National Astronomical Observatory of Japan, 2-21-1 Osawa, Mitaka, Tokyo 181-8588, Japan}

\author{Masayuki Akiyama}
\affiliation{Astronomical Institute, Tohoku University, 6-3 Aramaki, Aoba-ku, Sendai, Miyagi 980-8578, Japan}

\author{Yuichi Harikane}
\affil{National Astronomical Observatory of Japan, 2-21-1 Osawa, Mitaka, Tokyo 181-8588, Japan}

\author{Kohei Ichikawa}
\affiliation{Frontier Research Institute for Interdisciplinary Sciences, Tohoku University, Sendai, Miyagi 980-8578, Japan}
\affil{Astronomical Institute, Tohoku University, 6-3 Aramaki, Aoba-ku, Sendai, Miyagi 980-8578, Japan}

\author{Toshihiro Kawaguchi}
\affiliation{Department of Economics, Management and Information Science, Onomichi City University, Onomichi, Hiroshima 722-8506, Japan}

\author{Taiki Kawamuro}
\affiliation{National Astronomical Observatory of Japan, 2-21-1 Osawa, Mitaka, Tokyo 181-8588, Japan}

\author{Kotaro Kohno}
\affiliation{Institute of Astronomy, School of Science, The University of Tokyo, 2-21-1 Osawa, Mitaka, Tokyo 181-0015, Japan}
\affil{Research Center for the Early Universe, The University of Tokyo, 7-3-1 Hongo, Bunkyo, Tokyo 113-0033, Japan}

\author{Chien-Hsiu Lee}
\affiliation{National Optical Astronomy Observatory, 950 N. Cherry Ave., Tucson, AZ 85719, USA}

\author{Kianhong Lee}
\affiliation{Institute of Astronomy, School of Science, The University of Tokyo, 2-21-1 Osawa, Mitaka, Tokyo 181-0015, Japan}

\author{Yoshiki Matsuoka}
\affiliation{Research Center for Space and Cosmic Evolution, Ehime University, 2-5 Bunkyo-cho, Matsuyama, Ehime 790-8577, Japan}

\author{Mana Niida}
\affiliation{Graduate School of Science and Engineering, Ehime University, Bunkyo-cho, Matsuyama, Ehime 790-8577, Japan}

\author{Kazuyuki Ogura}
\affiliation{Faculty of Education, Bunkyo University, 3337, Minami-ogishima, Koshigaya, Saitama 343-8511, Japan}
\affiliation{Nishi-Harima Astronomical Observatory, Center for Astronomy, University of Hyogo, 407-2 Nishigaichi, Sayo, Hyogo 679-5313, Japan}

\author{Masafusa Onoue}
\affiliation{Max-Planck-Institut f\"ur Astronomie, K\"unigstuhl 17, D-69117 Heidelberg, Germany}

\author{Hisakazu Uchiyama}
\affiliation{National Astronomical Observatory of Japan, 2-21-1 Osawa, Mitaka, Tokyo 181-8588, Japan}

\begin{abstract}
We report a discovery of $z = 4.72$ radio galaxy, HSC J083913.17+011308.1,
by using the Lyman break technique 
with the Hyper Suprime-Cam Subaru Strategic Survey (HSC-SSP) catalog for VLA FIRST radio sources. 
The number of known high-$z$ radio galaxies (HzRGs) at $z > 3$ is quite small to constrain the evolution of HzRGs so far. 
The deep and wide-area optical survey by HSC-SSP enables us to apply the Lyman break technique to a large search for HzRGs. 
For an HzRG candidate among pre-selected $r$-band dropouts with a radio detection, 
a follow-up optical spectroscopy with GMOS/Gemini has been performed.
The obtained spectrum presents a clear Ly$\alpha$ emission line redshifted to $z=4.72$. 
The SED fitting analysis with the rest-frame UV and optical photometries suggests
the massive nature of this HzRG with $\log{M_*/M_{\sun}} = 11.4$.
The small equivalent width of Ly$\alpha$ and the moderately red UV colors 
indicate its dusty host galaxy, implying a chemically evolved and dusty system.
The radio spectral index does not meet a criterion for an ultra-steep spectrum: 
$\alpha^{325}_{1400}$ of $-1.1$ and $\alpha^{150}_{1400}$ of $-0.9$, 
demonstrating that the HSC-SSP survey compensates for a sub-population of HzRGs which 
are missed in surveys focusing on an ultra-steep spectral index.
\end{abstract}

\keywords{galaxies: active --- galaxies: high-redshift --- radio continuum: galaxies}


\section{Introduction}\label{sec:intro}

Observations and numerical studies suggest that
feedback effects from radio galaxies (RGs) can play a key role on the evolution of galaxies
by regulating star formation 
\citep[e.g.,][]{Croton06, Wagner11, Fabian12, Morganti13, McNamara05}.
It is important to describe the number densities of RGs along the cosmic time up to the early Universe for
understanding the evolution of galaxies and active galactic nuclei (AGNs). 
In particular, high-$z$ RGs (HzRGs) at $z > 2$ represent a key population to reveal 
the formation of massive galaxies and massive galaxy clusters \citep{MileyDeBreuck08},
because the stellar mass of HzRGs is typically as massive as $>10^{11} M_{\odot}$ even at $z > 3$ \citep{Seymour07, DeBreuck10}
and the gas metallicity of HzRGs at $z>3$  exceeds the solar metallicity \citep{Nagao06, Matsuoka09}.
The most massive galaxies could be rapidly formed before $z \sim 3$ \citep[e.g.,][]{Perez-Gonzalez08}.
The number density of RGs moderately increases up to $z \sim 1-2$ and
is believed to abruptly decline at the HzRG regime at $z = 2-3$ \citep{Dunlop90, Rigby15}.
At $z > 3$, the number density of HzRGs is not well constrained due to a dearth of samples,
except a small number of known radio-loud quasars \citep{Banados2015}.

HzRGs have been historically discovered by using a radio spectral index.
HzRGs are empirically known to show an ultra-steep spectrum (USS), 
which is commonly defined as a spectrum with a radio spectral index $\alpha$ of $<-1.3$
\citep{DeBreuck00, Saxena18a, Saxena19}.
The USS selection is based on an idea of a steepening radio spectrum with frequency and redshifting 
\citep[e.g.,][]{Carilli99, Morabito18}.
The technique using USS was successfully established and is an efficient 
method to find a lot of HzRGs \citep{DeBreuck00}.
The $z=5.19$ HzRG, TN J0924$-$2201, had been discovered with the USS technique 
and had been the most distant known HzRG record in the past two decades \citep{vanBreugel99}.
Recently, \citet{Saxena18b} broke the redshift record and discovered a $z=5.72$ HzRG, TGSS J1530+1049, using the USS technique.

However, a small number of $z>4$ HzRGs without USS has been discovered:
non-USS HzRGs at $z=4.42$ \citep[VLA J123642+621331,][]{Waddington99} and $z=4.88$ \citep[J163912.11+405236.5,][]{Jarvis09} were 
discovered out of radio-detected but very faint objects in optical and near-infrared, respectively.
The spectral indices of these two HzRGs are $-0.94$ between 1.4 GHz and 8.5 GHz 
and $-0.75$ between 325 MHz and 1.4 GHz, respectively. 
Their radio morphologies are compact and must be associated with an AGN.
\citet{Klamer06} indicated that no HzRGs selected based on USS show a curvature in gigahertz spectra.
These facts suggest the known HzRGs are a tip of the iceberg. 
Therefore the true spectral index distribution in HzRGs is unclear.
The present biased sample of HzRGs toward USS could lead to misunderstandings of 
the number density of HzRGs and underlying radio luminosity functions \citep{Jarvis00, Jarvis01, Yuan16, Yuan17}.

Finding HzRGs without relying on the USS technique is now possible 
by using the Lyman break technique on deep optical imaging \citep[e.g.,][]{Steidel96}.
Great advances of instruments capable deep imaging surveys in wide fields enable us 
to identify very faint and rare objects such as HzRGs.
A search for HzRGs using the Lyman break technique will provide 
a uniform HzRG sample without biases of radio spectral indices. 
Hyper Suprime-Cam Subaru Strategic Survey 
\citep[HSC-SSP,][]{Miyazaki18, Aihara18_survey, Komiyama18, Kawanomoto18, Furusawa18}
is one of the most suitable programs for this purpose.
The Lyman break technique for HSC-SSP photometric objects was
successfully applied to search for high-$z$ galaxies \citep{Ono18, Harikane18, Toshikawa18} 
and high-$z$ quasars \citep[e.g.,][]{Akiyama18, Matsuoka16, Matsuoka18a, Matsuoka18b, Matsuoka19}
In this paper, we report a discovery of $z$=4.7 HzRG, HSC J083913.17+011308.1 (hereafter HSC J0839+0113), 
found out of a pilot survey sample using Subaru HSC and an archival radio catalog. 
This survey is part of ongoing project, WERGS (Wide and Deep Exploration of RGs with Subaru HSC; \citealt{Yamashita18}). 
This paper is the third in a publication series of WERGS, which follows \citet{Yamashita18} and \citet{Toba19} and precedes Ichikawa et al.\ submitted.
Throughout this paper, 
all magnitudes are presented in the AB system \citep{Oke83} and
are corrected for Galactic extinction \citep{Schlegel98}.
We use the CModel photometry for HSC data \citep{Abazajian04}.
We use a flat $\Lambda$CDM cosmological model with $H_0 = 70$~km~s$^{-1}$~Mpc$^{-1}$, and $\Omega_M = 0.30$.
Using this cosmology, at $z=4.7$ the age of the Universe is 1.2~Gyr and 1$\arcsec$ corresponds to 6.5~kpc.

\section{Selection}
We utilized the 1.4 GHz radio continuum catalog of the Faint Images of the Radio Sky at Twenty-cm survey 
\citep[FIRST,][]{Becker95} with Very Large Array (VLA).
FIRST surveyed the area of the Sloan Digital Sky Survey \citep[SDSS][]{York00} and 
thus entirely covers the HSC-SSP survey fields
with a spatial resolution of 5\farcs4.
We only used radio sources with a peak flux of greater than 1~mJy and 
a side lobe probability\footnote{The side lobe probability is a measure of a likelihood that a source is a spurious detection near a bright source. } 
of less than 0.05 in the final release catalog \citep{Helfand15}.

In order to identify optical counterparts of FIRST sources,
we used a forced photometry catalog in the wide-layer of the internally released
version, S16A Wide2, of HSC-SSP \citep{Aihara18_DR1}.
The forced photometries were performed for each image using position and shape parameters 
determined in a reference band. 
The order of the priority bands as a reference band is $irzyg$ 
according to its signal-to-noise ratio (S/N) approximately
\citep[see][for more details]{Bosch18}.
Only unique object that has been deblended and photometried based on $i$-band 
was selected by flags, 
\verb|detect_is_primary| and \verb|merge_measurement_i|, in the HSC-SSP database.
Further, we imposed the following criteria on objects: 
in $i$, $r$, and $z$-bands,
an object is not affected by cosmic rays (\verb|flags_pixel_cr_center|),
not saturated (\verb|flags_pixel_saturated_center|),
and not at the edge of a CCD or a coadded image (\verb|flags_pixel_edge| for removing); 
in $i$ and $z$-bands, 
there are no problems in photometries (\verb|cmodel_flux_flags|, \verb|centroid_sdss_flags|);
the number of visits at an object position (\verb|countinputs|) is greater than or equal 
to 4 (6) for the g and $r$ ($i$ and $z$) bands to ensure a certain observing depth. 
We have cross-matched between the FIRST sources and the HSC-SSP sources with a search radius of $1\arcsec$.
This search radius was set to balance between contamination and completeness,
where the estimated contamination by chance coincidence is 14\% (see \citealt{Yamashita18} for the details).

\begin{figure}[!t]
\epsscale{1.1}
\begin{center}
\plotone{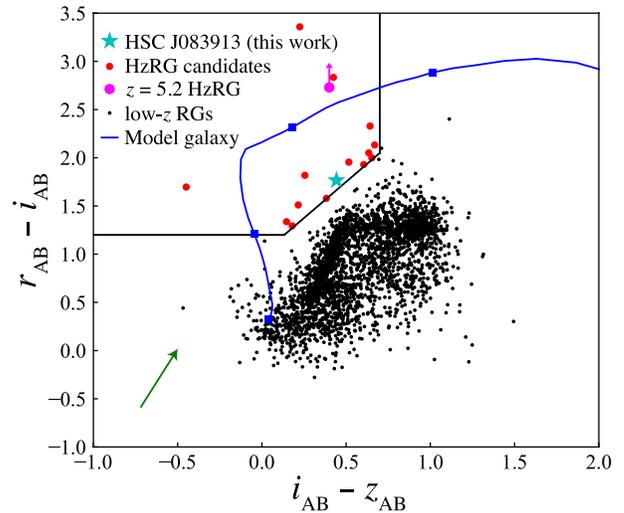}
\caption{
Color-color diagram for selecting HzRG candidates at $z\sim 5$.
The candidates are denoted by the red circles 
and are inside the color selection criteria (black lines, \citealt{Ono18}). 
One candidate is out of the y-axis range. 
The cyan star represents HSC J0839+0113. 
The magenta circle denotes the colors of the known HzRG at $z=5.2$, where
used filters were slightly different from HSC's ones and $y$-axis is a lower limit \citep{Overzier06}.
The blue line indicates the color track of a star-forming galaxy model produced with a stellar synthesis code of \citet{BC03}.
The model assumes an instantaneous-burst model with an age of 25 Myr, the solar metallicity, 
the IGM absorption of \citet{Madau95}, and 
Lyman$\alpha$ emission with a rest-frame equivalent width of 1180 \AA\ and 
an observed frame FWHM of 1530 km s$^{-1}$ (the most large equivalent width case in $z>2$ HzRGs of \citealt{Roettgering97}).
The squares on the track mark redshifts of 4.0, 4.5, 5.0, and 5.5.
The black dots show low-$z$ RGs in the WERGS sample \citep{Yamashita18}.
The green vector shows the extinction vector of $A_{\rm V} = 1$ \citep{Calzetti00}.
\label{fig:riz}}
\end{center}
\end{figure}

\begin{figure*}
\epsscale{1.15}
\begin{center}
\plotone{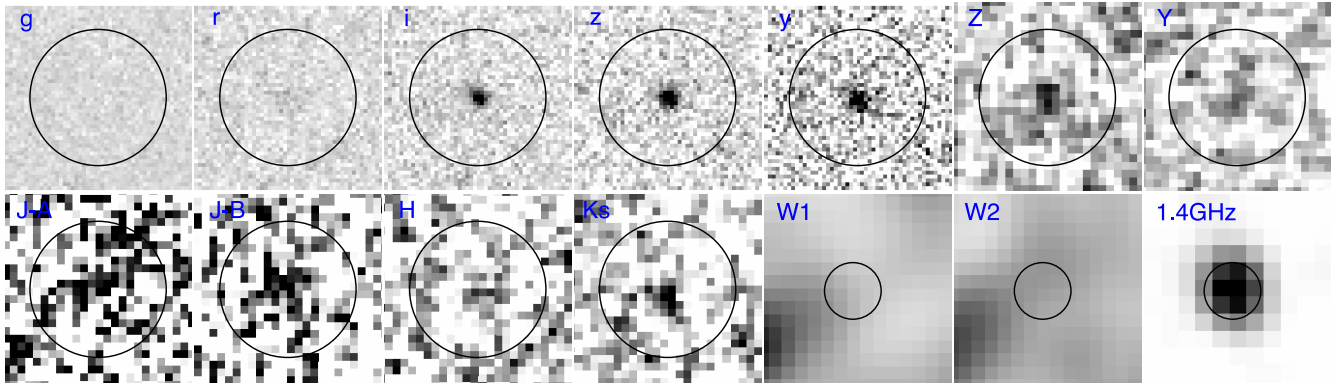}
\caption{
The postage stamp images of HSC J0839+0113. 
From top left to bottom right, 
the Subaru HSC $grizy$ and VIKING $ZYJHKs$ images shown 
with a width of 8\arcsec, 
and subsequently the {\it WISE} $W1W2$ and FIRST 1.4~GHz images
with a width of 20\arcsec.
Two frames of VIKING $J$-band, denoted by $J$-A and $J$-B, respectively, are shown
and the results of SED fitting with each $J$-band photometries are discussed in Section \ref{sec:hostgalaxy}.
North is upward and east is to the left.
There are no other optical/near-infrared sources within the FIRST detection region (FWHM $\sim 6\arcsec$).
The infrared counterparts in both $W1$ and $W2$ are not detected. 
Circles with an equivalent diameter of 6\arcsec to the apparent size in the FIRST image
serve as a guide to the eye.
\label{fig:img}}
\end{center}
\end{figure*}

Out of 4725 matched sources, 
16 sources were selected as $r$-dropouts with S/N $> 5$ in $z$-band, 
which meet the following color criteria which are the same ones as those of \citet{Ono18}: 
\begin{eqnarray}
r_{\rm AB} - i_{\rm AB} &>& 1.2 \\
i_{\rm AB} - z_{\rm AB} &<& 0.7 \\
r_{\rm AB} - i_{\rm AB} &>& 1.5\,(i_{\rm AB}-z_{\rm AB}) + 1.0
\end{eqnarray}
The objects further should be not detected in $g$ band or $g_{\rm AB} > r_{\rm AB}$.
These $r$-dropouts are candidates of HzRGs at $z \sim 5$ (Figure \ref{fig:riz}).
HSC J0839+0113 was selected out of them according to the brightness ($z_{\rm AB}<25$) and
the visibility for a follow-up observation with GMOS/Gemini (see the next section). 
HSC J0839+0113 has a FIRST 1.4 GHz flux of 7.17$\pm 0.14$ mJy (see also Table \ref{tab:properties}).
Its HSC images and FIRST radio continuum are displayed in Figure \ref{fig:img}.
The obvious dropout at $r$-band is seen. 
The apparent size in the FIRST image is approximately a point source (FWHM $= 6\farcs25 \times 5\farcs30$), 
as seen in high-$z$ radio sources.
There are no other HSC sources within the FIRST beam.

\section{Spectroscopic Redshift and Photometric Data}
\subsection{Spectroscopic Follow-up Observation}
A long slit spectroscopy was performed for HSC J0839+0113 using GMOS mounted on the Gemini South telescope 
on 2018, February 21 (Program ID: GS-2017B-FT-17, PI: T. Yamashita),
in order to obtain its spectroscopic redshift.
We used the OG515 filter and the R400-G5305 grating blazed at 8300 \AA.
The slit width was set to be 1\farcs0.
The spectral resolution is approximately 8 \AA, which corresponds to $290$~km~s$^{-1}$. 
The spectral coverage was typically 6700 -- 10000 \AA.
The position angle was set to be 32.56 degree East of North.
A total of $7 \times 1030$ seconds exposures were obtained. 
Bias frames, flat-fields, CuAr arcs, and standard star Hiltner~600, 
were also taken for calibration.
The seeing of the observation was 0\farcs99.

Data reduction was carried out using the Gemini/GMOS IRAF package.
Object frames were bias-subtracted and flat-fielded.
The wavelength calibration for the object frames was performed using reduced arc frames.
After removing cosmic rays, subtracting sky lines, and flux calibration using
the spectrophotometric standard star (Hiltner 600),
the object frames were combined with median stacking.
A 1D spectrum was extracted with an aperture of 1\farcs9.

\begin{figure*}
\epsscale{1.1}
\begin{center}
\plotone{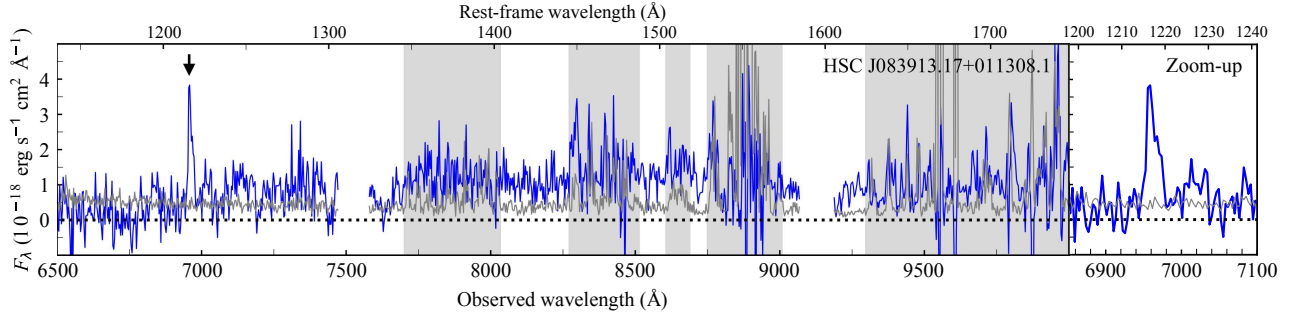}
\caption{
The GMOS spectrum of HSC J0839+0113.
The redshifted Ly$\alpha$ to $z = 4.723$ is shown at 6957 \AA\ in the observed-frame (black arrow).
Ly$\alpha$ break is also detected at the bluer part than the Ly$\alpha$ line.
The full spectrum (blue thin line) and the enlarged spectrum around the Ly$\alpha$ 
emission line (blue thick line) are shown.
A noise spectrum (gray line) serves as a guide to the expected noise. 
The wavelength regime strongly affected by sky lines are shaded.
The missing data points are due to CCD gaps.
\label{fig:spec}}
\end{center}
\end{figure*}

Ly$\alpha$ $\lambda1216$ emission line is significantly detected at a peak of 6957 \AA\
in the spectrum (Figure \ref{fig:spec}).
The redshifted Ly$\alpha$ corresponds to a redshift of $4.723 \pm 0.001$.
The characteristic asymmetric profile of this emission line
(see the right panel of Figure \ref{fig:spec})
and the $r$-dropout of this object credibly support that
this line is the redshifted Ly$\alpha$, 
although any other lines were not identified in this spectrum.
Continuum is also detected at the redder part than the Ly$\alpha$ line
while is not detected in the bluer part, which is consistent with Ly$\alpha$ break at this redshift.
The 2D spectrum shows an extended Ly$\alpha$ of 1\farcs3, which corresponds to 8.4~kpc.
This extended Ly$\alpha$ is approximately equivalent to a deconvolved size of 7.6~kpc in the 1.4~GHz image
produced by the FIRST survey project \citep{Helfand15}.

\subsection{SED Fitting Analysis}
We estimate a stellar mass of this HzRG by a fitting analysis of its spectral 
energy distribution (SED). 
Photometric data of VIKING \citep{Edge13} and ALLWISE \citep{Cutri14} are available 
for this object, as well as HSC-SSP (see Figure \ref{fig:img} for images of all bands). 
Because no entry of this object is in the VIKING DR3 source catalog likely due to its low S/N, 
we performed aperture photometries on archive VIKING images of all $ZYJHKs$-bands.
The VIKING images are convolved to match their PSF to one of $Z$-band, 
after sky backgrounds are subtracted.
The $2\farcs5$ aperture is used for the object photometries. 
This aperture radius at the object position encloses an equivalent flux of a $i$-band image 
convolved to $Z$-band to the CModel $i$-band flux. 
Therefore the $2\farcs5$ aperture photometries can measure a total flux of the object.
The errors in the photometries are estimated from variations of aperture photometries 
at object-free regions.
Only in $J$-band, two archive image frames are available for this object. 
We note that two photometry results of these frames are significantly different from each other. 
This large difference is likely due to an effect from large-scale variations of sky backgrounds on the low S/N images.
Thus, we use a weighted mean of the two photometries for the SED fitting analysis. 
Finally, we obtain 23.81 (0.03), 23.1 (0.4), 22.6 (0.5), $>22.4$, 
and 22.4 (0.5) in $ZYJHKs$ bands, respectively. 
The parentheses represent $1\sigma$ errors. 
For a non-detection in $H$-band, the $3 \sigma$ limit is provided.
In ALLWISE, there are no entries in the catalog of both bands of $W1$ and $W2$. 
Even in the images, the object is not detected in both bands, 
after a possible contamination from an adjacent object is considered
(a source at the south-east from HSC J0839+0113 in Figure \ref{fig:img}).
We thus put  $3 \sigma$ upper limits of 32~$\mu$Jy and 43~$\mu$Jy for $W1$ and $W2$, respectively.

SED fitting is performed with these optical-to-mid-infrared photometries including upper limits 
using the CIGALE SED fitting code \citep{Burgarella05, Boquien19}. 
We basically follow a SED fitting manner of \citet{Toba19}, 
who applied the CIGALE SED fitting for HSC-selected radio galaxies,
but we use somewhat smaller steps of free parameters than \citet{Toba19}
and do not include models of infrared dust emission, radio synchrotron emission, 
and AGN emission.
In our SED fitting analysis, 
the star formation history of two exponential decreasing star formation with different 
$e$-folding times is assumed. 
A stellar template of \citet{BC03} with an initial mass function of \citet{Chabrier03} is adopted. 
Dust attenuation is modeled by the dust extinction law of \citet{Calzetti00}.
The SED fitting with CIGALE also handle fluxes with upper limits by introducing the error functions for computing $\chi^2$
\citep[see][]{Boquien19}.

\begin{figure}[t]
\epsscale{1.1}
\begin{center}
\plotone{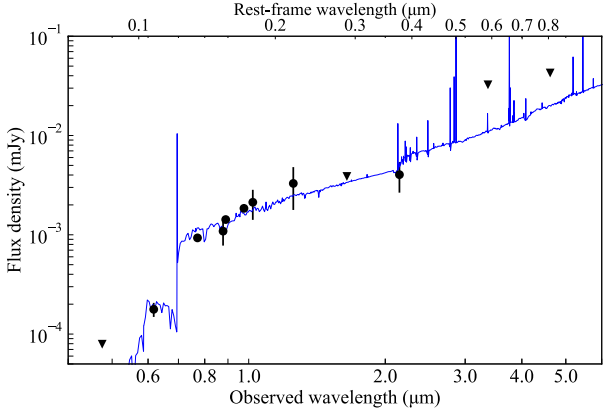}
\caption{
The SED of HSC J0839+0113 and its best fit models. 
The plotted data are photometries of $grizy$ (HSC), $ZYJHKs$ (VIKING), and $W1W2$ ({\it WISE}).
The circles and triangles denote detections with $1\sigma$ error bars 
and non-detection ($3\sigma$ upper limits), respectively. 
\label{fig:sed}}
\end{center}
\end{figure}

The best-fit SED model is presented in Figure \ref{fig:sed}.
The reduced $\chi^{2}$ of the best-fit is 0.46 with a degree of freedom of 10.
The bayesian-estimated stellar mass is $\log{M_*/M_{\sun} = 11.43^{+0.22}_{-0.46}}$, 
where the super/subscripts are $1\sigma$ errors.
The mass-weighted age is 230 (160) Myr. 
We cross-checked the stellar mass with that estimated with the Mizuki SED template-fitting code \citep{Tanaka15}, 
where the Mizuki fitting was performed for all the photometry data of this object.
The estimated stellar mass is $11.66_{-0.26}^{+0.24}$.
The Mizuki result is consistent with that of CIGALE within the $1\sigma$ errors. 
We do not provide the star formation rate (SFR) from this SED fitting analysis 
because the SFR is not constrained 
due to the less number of photometry data in the observed-frame infrared wavelength.
The results are further discussed in Section \ref{sec:hostgalaxy}.

\begin{table}[t!]
\begin{center}
\caption{Properties of HSC J$083913.17+011308.1$} 
\label{tab:properties}
\begin{tabular}{lll}
\tablewidth{0pt}
\hline
\hline
\decimals
R.A, Decl. (J2000)&  08$^{\mathrm h}$39\arcmin13\farcs17, $+$01\arcdeg13\arcmin08\farcs1   \\
Redshift &  4.723 (0.001)\\
\hline
$r_{\rm AB}$, $i_{\rm AB}$,   &  25.66 (0.18), 23.89 (0.03) \\
$z_{\rm AB}$, $y_{\rm AB}$   & 23.45 (0.04), 23.23 (0.06) \\
$Z_{\rm AB}$, $Y_{\rm AB}$, $J_{\rm AB}$  & 23.81 (0.03), 23.1 (0.4), 22.6 (0.5) \\
$H_{\rm AB}$, $Ks_{\rm AB}$         & $>22.4$\tablenotemark{a}, 22.4 (0.5) \\    
FIRST $S_{1.4\rm GHz}$
           & 7.17 (0.14) mJy \\
GMRT $S_{325\rm MHz}$
           & 36.8 (2.0) mJy \\
TGSS $S_{150\rm MHz}$
           & 54.5 (6.6) mJy\\
$L_{1.4\rm GHz}$
           &  1.63 (0.03) $\times 10^{27}$ W Hz$^{-1}$\\
$L_{325\rm MHz}$
           & 8.38 (0.46) $\times 10^{27}$ W Hz$^{-1}$\\
$L_{150\rm MHz}$
           & 12.4 (1.5) $\times 10^{27}$ W Hz$^{-1}$ \\
$\alpha^{325}_{1400}$ 
           & $-1.12$  (0.02)\\
$\alpha^{150}_{1400}$ 
           & $-0.91$ (0.02) \\
\hline
Ly$\alpha$ flux
         & 4.3 (0.4) $\times 10^{-17}$ erg s$^{-1}$ cm$^{-2}$\\
Ly$\alpha$ luminosity
         & 9.80 (0.10) $\times 10^{42}$ erg s$^{-1}$\\      
Ly$\alpha$ FWHM
         & 660 (90) km s$^{-1}$\\
Ly$\alpha$ EW$_{\rm rest}$
		 & 9.1 (1.6) \AA \\
\hline
$\log{M_{*}/M_{\odot}}$\tablenotemark{b} & 11.43 ($^{+0.22}_{-0.46}$) \\
\hline
\end{tabular}
\end{center}
\tablecomments{1$\sigma$ errors are shown in parentheses.}
\tablenotetext{a}{3$\sigma$ upper limit.}
\tablenotetext{b}{The stellar mass is derived from SED fitting.}
\end{table}

\section{Discussion: the Nature of HSC J0839+0113}
\subsection{Radio Properties}

HSC J0839+0113 was selected from Lyman break galaxies (LBGs) with a radio detection of FIRST. 
Because any criteria on a radio spectral index are not imposed to the selection, 
discussing its radio properties is relevant. 
Radio photometric data of HSC J0839+0113 are available in the FIRST 1.4~GHz catalog \citep{Helfand15}, 
the 325 MHz GMRT survey catalog in the {\it Herschel}-ATLAS/GAMA field \citep{Mauch13}, 
and the 150 MHz GMRT TGSS Alternative Data Release \citep{Intema2017}.

The spectral indices $\alpha$ ($S_{\nu} \propto \nu^{\alpha}$) are calculated to be 
$\alpha^{150}_{1400} = -0.91$ (0.02) between 150 MHz and 1.4 GHz, 
and $\alpha^{325}_{1400} = -1.12$ (0.02) between 325 MHz and 1.4 GHz, respectively.
Alternative spectral index of $\alpha^{150}_{1400}$ is $-0.84$ in the spectral index catalog by 
\citet{deGasperin2018} who re-processed the TGSS and NVSS images. 
These indices do not meet the criterion defining ultra-steep spectral index, 
$\alpha < -1.3$ \citep{DeBreuck00, Saxena18a}.

\begin{figure}[t]
\epsscale{1}
\begin{center}
\plotone{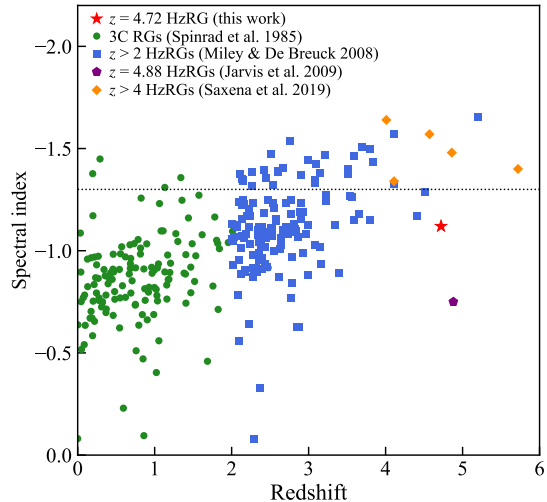}
\caption{
Radio spectral index ($\alpha$; $S_{\nu} \propto \nu^{\alpha}$) 
as a function of redshift of RGs.
The red star indicates the HSC J0839+0113 in this paper.
The known RGs taken from literature are also plotted:
3CR \citep[green circles,][]{Spinrad85}, 
the HzRG sample at $z>2$ \citep[blue squares,][]{MileyDeBreuck08}, 
the $z = 4.88$ non-USS HzRG \citep[purple pentagon,][]{Jarvis09}, 
and the five recently discovered HzRGs at $z>4$ \citep[orange diamonds,][]{Saxena19}.
The commonly used criterion of $\alpha$ for USS selection is represented by a dotted line.
\label{fig:indexz}}
\end{center}
\end{figure}

In Figure \ref{fig:indexz}, the spectral index between 325 MHz and 1.4 GHz is shown
as a function of redshift together with other known HzRGs. 
The spectral indices of the other RGs and HzRGs are taken from literature:
3CR \citep{Spinrad85}, 
the HzRG sample at $z>2$ \citep{MileyDeBreuck08}, 
the $z = 4.88$ non-USS HzRG with $\alpha = -0.75$ \citep{Jarvis09}, 
and the five recently discovered HzRGs at $z>4$ \citep[$\alpha$ of from $-1.34$ to $-1.64$,][]{Saxena19}.
The spectral indices of the literature samples are calculated 
between WENSS 325 MHz \citep{Rengelink97} and NVSS 1.4 GHz \citep{Condon98} or 
between TEXAS 365 MHz \citep{Douglas96} and NVSS 1.4 GHz 
except for Saxena's HzRGs. 
The spectral indices for $z > 4$ Saxena's HzRGs are calculated between 
VLA 370 MHz and 1.4 GHz \citep{Saxena18a}
except for the most distant known HzRG at $z=5.72$ (TGSS J1530+1049) with 
a spectral index of $-1.4$ between TGSS 150 MHz and VLA 1.4 GHz \citep{Saxena18b}. 
We can see the $\alpha - z$ correlation at all the redshift range.
At $z > 4$, $z=4.88$ HzRG (J163912.11+405236.5) shows the flattest spectral index. 
This object has been discovered out of faint mid-infrared sources with a FIRST detection
in the {\it Spitzer}-SWIRE deep field, not by using the USS technique \citep{Jarvis09}. 
This is an evident example illustrating non-USS HzRGs at $z>4$.
VLA J1236421+621331, another non-USS HzRG at $z=4.42$ \citep{Waddington99}, is not plotted here
because the low frequency photometry data is not available. 
HSC J0839+0113 is located below the $\alpha = -1.3$ selection line (the dotted line in Figure \ref{fig:indexz}), 
following the $z=4.88$ HzRG at $z>4$.
HSC J0839+0113 would be never discovered by using the USS technique. 
This discovery of the non-USS HzRG at $z\sim 5$, 
along with the previously reported non-USS HzRGs at $z\sim 4-5$ \citep{Waddington99, Jarvis09},
suggests that a subset of HzRGs are missed in many USS-based surveys.

The rest-frame radio luminosities of HSC J0839+0113
are $\log{L_{1.4\rm GHz}}$ of 27.2~W~Hz$^{-1}$ and $\log{L_{150\rm MHz}}$ of 28.1~W~Hz$^{-1}$,
assuming a mean spectral index, $\alpha = -1.0$, between $\alpha^{325}_{1400}$ and $\alpha^{150}_{1400}$.
Owing to the relatively flat spectrum, the radio luminosities at low frequency
is one order of magnitude smaller than those of the previously reported USS-selected HzRGs at $z\sim 4$ -- 5, 
and is a factor of 2--4 smaller than those of the recently reported USS-selected HzRGs at $z\sim 4$ -- 5 \citep{Saxena19}.

\subsection{Host galaxy}\label{sec:hostgalaxy}
HzRGs are often involved with massive hosts \citep[$>10^{11} M_{\odot}$,][]{Seymour07, DeBreuck10}.
We can examine whether HSC J0839+0113 that was selected among LBGs 
follows the trend or not, using the SED fitting result.
The obtained stellar mass of HSC J0839+0113 is $\log{M_*/M_{\sun}} = 11.4$ and
is as massive as other known HzRGs \citep{Seymour07, DeBreuck10}. 
This massive stellar mass of the host galaxy is supported from a fact that
the apparent $Ks$ magnitude of 20.6 Vega mag well follows the $K-z$ correlation for HzRGs, 
where the correlation is established because of the characteristic massive stellar mass 
of known HzRGs \citep[e.g.,][]{Rocca-Volmerange04}. 
HSC J0839+0113, which is a LBG with a radio-AGN, is one of the most massive galaxies among $z\sim 5$ LBGs, 
which typically have $\log{M_*/M_{\sun}} = 8 - 11$ \citep{Yabe09}.
Such a large stellar mass at $z=4.72$ (at an age of the universe of 1.2 Gyr)
suggests that the stellar mass had been rapidly built up through intense star formation.
Since HSC J0839+0113 is already very massive at the observed redshift,
the star formation should either have been quenched by now or be quenched shortly, 
and then this HzRG could evolve into a giant elliptical galaxy seen at the present day.

\subsection{Ly$\alpha$ Emission}
The measured Ly$\alpha$ line flux and luminosity of HSC J0839+0113 
are $4.3 \times 10^{-17}$ erg~s$^{-1}$~cm$^{-2}$ and
$9.8\times 10^{42}$ erg~s$^{-1}$, respectively. 
The underlying continuum is estimated from the continuum of 7000--7300 \AA\ 
in the spectrum, assuming a flat continuum. 
Ly$\alpha$ emission is not corrected for extinction.
The Ly$\alpha$ luminosity is consistent with 
those of other $z \sim 4-5$ HzRGs \citep{vanBreugel99, Jarvis09, Kopylov06, Saxena19}.
The Ly$\alpha$ FWHM is derived with
a single Gaussian fitting and is corrected for the instrumental broadening effect. 
The derived FWHM is 660 km~s$^{-1}$. 
The FWHM is smaller than those of the known HzRGs at $z > 2$, 
which have typically $>1000$ km s$^{-1}$.
This narrow FWHM of Ly$\alpha$ could suggest that 
a dominant ionization source of Ly$\alpha$ of HSC J0839+0113 is 
neither a broad emission line AGN nor jet-induced shock but star formation or a weak AGN. 
\citet{Saxena19} argue that the dominant source of faint HzRGs 
with low Ly$\alpha$ luminosities ($\sim 10^{43}$ erg~s$^{-1}$) and narrow FWHMs ($<600$ km s$^{-1}$) at $z>4$ 
could be star formation or a weak AGN by compared with Lyman alpha emitters. 
Our result is not inconsistent with that.

Equivalent width EW$_{\rm rest}$ of Ly$\alpha$ at the rest-frame 
is calculated to be 9 (1.6)~\AA\ from the obtained spectrum. 
The EW$_{\rm rest}$ of HSC J0839+0113 is much smaller than those of the known HzRGs. 
The known HzRGs at $2<z<4$ show large EW$_{\rm rest}$ of up to $\sim$700 \AA. 
For almost all of the HzRGs at $z>4$, continua are not detected and 
the upper limits of EW$_{\rm rest}$ are measured to be $>40 - 180$ \AA\ \citep{DeBreuck00_StatSpec, Waddington99, Saxena18b}. 
The small EW$_{\rm rest}$ of HSC J0839+0113 is probably due to a mixture of its ionization source and optical brightness.
UV emission lines of HzRGs are generally explained by photoionization by a powerful AGN and additionally jet-induced shock \citep[e.g.,][]{DeBreuck00_StatSpec}. 
As discussed above, however, in HSC J0839+0113 the contributions from a powerful AGN or shock must be insignificant.
The different ionization source in HSC J0839+0113, therefore, could result in the weak Ly$\alpha$ emission. 
In addition to the ionization source, the relative bright optical continuum of HSC J0839+0113 ($z_{\rm AB} = 23.5$) 
also likely causes the small EW$_{\rm rest}$.
The rest-frame UV continuum emission of HSC J0839+0113 is detected both in HSC and the Gemini spectroscopy,
while $z>4$ HzRGs are not detected ($>23$ mag).

A small EW$_{\rm rest}$ is observed as the Ly$\alpha$ deficiency in luminous LBGs \citep{Ando06}
and is attributed to a dusty absorption in a star-forming galaxy. 
The obtained small EW$_{\rm rest}$ and the relatively optical brightness indicate its dusty system associated with HSC J0839+0113.
This is consistent with the migration on the $riz$ diagram in Figure \ref{fig:riz} from the model colors of a $z\sim 4.7$ LBG 
to redder colors which suggests this dusty system. 
The colors of HSC J0839+0113 cannot be explained by galaxy models with low metallicity, 
although a low metallicity galaxy at $z \sim 5.2$ could have a similar color to that of HSC J0839+0113.
These results support the abundant dust contents in HSC J0839+0113 and suggest the chemically evolved host. 
Submillimeter continua are detected in approximately half of HzRGs at $2.5 < z < 4$ \citep{Reuland04}, 
indicating the presence of dust. 
Only two $z>4$ HzRGs (TN J1338-1942 at $z=4.11$ and TN J0924-2201 at $z=5.19$)
were observed with (sub)millimeter telescopes and their continua were detected \citep{DeBreuck0004, Falkendal19}. 
Although these two cases suggest the presence of cold dust there
despite their large Ly$\alpha$ EW$_{\rm rest}$ of 200 \AA\ and $>180$ \AA\ respectively, 
however, their Ly$\alpha$ emission are likely powered by a broad emission line AGN or jet-induced shock 
because of their large Ly$\alpha$ velocity widths of 1000 km s$^{-1}$ and 1500 km s$^{-1}$.

\section{Summary}

This paper reports the discovery of HSC J0839+0113, a $z=4.72$ HzRG, in the on-going program to search for
HzRGs as part of the Wide and Deep Exploration of Radio Galaxies with Subaru HSC (WERGS) project. 
HSC J0839+0113 was selected as a HzRG candidate from the HSC data using the Lyman break technique. 
The redshift was determined with Ly$\alpha$ emission line in the obtained Gemini/GMOS spectrum.
From the analyses of the GMOS spectrum and photometric data, we found:
\begin{itemize}
\item  
By the SED fitting with the rest-frame UV-optical photometries, 
the massive stellar mass of the host galaxy was estimated to 
be $\log{M_*/M_{\sun} = 11.4}$ and is consistent with other known HzRGs 
but marks one of the most massive galaxies among LBGs at $z=5$.
\item   
The Ly$\alpha$ line luminosity and FWHM are similar to those of other known $z\sim 4-5$ HzRGs.
The small $EW_{\rm rest}$ of HSC J0839+0113 is much different from HzRGs at $2<z<4$ 
but consistent with the Ly$\alpha$ deficiency in luminous LBGs, 
suggesting a dusty host galaxy of HSC J0839+0113. 
The rest-frame UV colors seen on the $riz$ color-color diagram is redder than 
those of the $z\sim 5$ LBG model and also supports the dusty and chemically evolved host galaxy.
\item 
HSC J0839+0113 has the relatively flat radio spectral index $\alpha^{325}_{1400}$ of $-1.1$ 
and thus does not belong to USS HzRGs.
\end{itemize}

The discovery of HSC J0839+0113 demonstrates the ability of HSC-SSP to find HzRGs without a radio spectral index. 
A HzRG sample established based on HSC-SSP 
compensates for a sub-population of HzRGs which are missed in USS selection surveys. 
Because the HzRG was found out of the pilot selection sample in the early release data of HSC-SSP,
the full dataset of HSC-SSP covering 1400 deg$^2$ has a potential for finding much more HzRGs at $z>3$, 
at which known HzRGs are poor in numbers.
HzRGs in our survey will provide a new knowledge on
the formation and evolution of massive galaxies and radio-AGNs at the early Universe.

\acknowledgments

We thank the anonymous referee for the comments that improved this paper. 
We would like to thank Hiroshi Nagai for useful discussions.

The Hyper Suprime-Cam (HSC) collaboration includes the astronomical communities of Japan and Taiwan, and Princeton University.  The HSC instrumentation and software were developed by the National Astronomical Observatory of Japan (NAOJ), the Kavli Institute for the Physics and Mathematics of the Universe (Kavli IPMU), the University of Tokyo, the High Energy Accelerator Research Organization (KEK), the Academia Sinica Institute for Astronomy and Astrophysics in Taiwan (ASIAA), and Princeton University.  Funding was contributed by the FIRST program from the Japanese Cabinet Office, the Ministry of Education, Culture, Sports, Science and Technology (MEXT), the Japan Society for the Promotion of Science (JSPS), Japan Science and Technology Agency  (JST), the Toray Science  Foundation, NAOJ, Kavli IPMU, KEK, ASIAA, and Princeton University.

This paper makes use of software developed for the Large Synoptic Survey Telescope \citep{Juric17, Ivezic19}. We thank the LSST Project for making their code available as free software at  http://dm.lsst.org.

This paper is based in part on data collected at the Subaru Telescope and retrieved from the HSC data archive system, which is operated by Subaru Telescope and Astronomy Data Center (ADC) at NAOJ. Data analysis was in part carried out with the cooperation of Center for Computational Astrophysics (CfCA), NAOJ. 

The Pan-STARRS1 Surveys (PS1, \citealt{Chambers16, Schlafly12, Tonry12, Magnier13}) and the PS1 public science archive have been made possible through contributions by the Institute for Astronomy, the University of Hawaii, the Pan-STARRS Project Office, the Max Planck Society and its participating institutes, the Max Planck Institute for Astronomy, Heidelberg, and the Max Planck Institute for Extraterrestrial Physics, Garching, The Johns Hopkins University, Durham University, the University of Edinburgh, the Queen’s University Belfast, the Harvard-Smithsonian Center for Astrophysics, the Las Cumbres Observatory Global Telescope Network Incorporated, the National Central University of Taiwan, the Space Telescope Science Institute, the National Aeronautics and Space Administration under grant No. NNX08AR22G issued through the Planetary Science Division of the NASA Science Mission Directorate, the National Science Foundation grant No. AST-1238877, the University of Maryland, Eotvos Lorand University (ELTE), the Los Alamos National Laboratory, and the Gordon and Betty Moore Foundation.

This study is based on observations obtained at the Gemini Observatory
via the time exchange program between Gemini and the Subaru Telescope, 
processed using the Gemini IRAF package.
The Gemini Observatory is operated by the Association of Universities for Research in Astronomy, Inc., 
under a cooperative agreement with the NSF on behalf of the Gemini partnership: 
the National Science Foundation (United States), National Research Council (Canada), 
CONICYT (Chile), Ministerio de Ciencia, Tecnolog\'{i}a e Innovaci\'{o}n Productiva (Argentina), 
Minist\'{e}rio da Ci\^{e}ncia, Tecnologia e Inova\c{c}\~{a}o (Brazil), and 
Korea Astronomy and Space Science Institute (Republic of Korea).

The National Radio Astronomy Observatory is a facility of the National Science 
Foundation operated under cooperative agreement by Associated Universities, Inc.

GMRT is run by the National Centre for Radio Astrophysics
of the Tata Institute of Fundamental Research.

This research has made use of the VizieR catalogue access tool, CDS, Strasbourg, France.

This work is financially supported by the Japan Society for the Promotion of Science (JSPS) KAKENHI grants: Nos.\
16H01101 (TN), 16H03958 (TN), 17H01114 (TN \& YO), 17H04830 (YM), 18K13584 (KI), 
18J01050 (YT), and 19K14759 (YT).
YM acknowledges the support from the Mitsubishi Foundation 
grant No.\ 30140.

\vspace{5mm}
\facilities{Subaru (HSC), Gemini:South (GMOS), VLA, GMRT, }

\software{Python, 
astropy, 
Gemini IRAF,
CIGALE}





\end{document}